\documentclass[%
 reprint,
 amsmath,amssymb,
 aps,
]{revtex4-2}

\usepackage{graphicx}
\usepackage{dcolumn}
\usepackage{bm}
\usepackage{ulem}

\usepackage{xcolor}

\begin{document}

\preprint{APS/123-QED}

\title{Unraveling Fano noise and partial charge collection effect \\ in X-ray spectra below 1 keV} 

\author{Dario Rodrigues}
\affiliation{\normalsize\it Universidad de Buenos Aires, Facultad de Ciencias Exactas y Naturales, Departamento de Física. Buenos Aires, Argentina.}
\affiliation{\normalsize\it CONICET - Universidad de Buenos Aires, Instituto de Física de Buenos Aires (IFIBA). Buenos Aires, Argentina}

\author{Mariano Cababie}
\affiliation{\normalsize\it Universidad de Buenos Aires, Facultad de Ciencias Exactas y Naturales, Departamento de Física. Buenos Aires, Argentina.}

\author{Ignacio Gomez Florenciano}
\affiliation{\normalsize\it Universidad de Buenos Aires, Facultad de Ciencias Exactas y Naturales, Departamento de Física. Buenos Aires, Argentina.}

\author{Agustina Magnoni}
\affiliation{\normalsize\it Universidad de Buenos Aires, Facultad de Ciencias Exactas y Naturales, Departamento de Física. Buenos Aires, Argentina.}
\affiliation{\normalsize\it Laboratorio de Óptica Cuántica, DEILAP, UNIDEF (CITEDEF-CONICET), Buenos Aires, Argentina}

\author{Ana Botti}
\affiliation{\normalsize\it 
Fermi National Accelerator Laboratory, PO Box 500, Batavia IL, 60510}

\author{Juan Estrada}
\affiliation{\normalsize\it 
Fermi National Accelerator Laboratory, PO Box 500, Batavia IL, 60510}

\author{Guillermo Fernandez-Moroni}
\affiliation{\normalsize\it 
Fermi National Accelerator Laboratory, PO Box 500, Batavia IL, 60510}

\author{Javier Tiffenberg}
\affiliation{\normalsize\it 
Fermi National Accelerator Laboratory, PO Box 500, Batavia IL, 60510}

\author{Sho Uemura}
\affiliation{\normalsize\it 
Fermi National Accelerator Laboratory, PO Box 500, Batavia IL, 60510}

\date{\today}

\hfill{FERMILAB-PUB-23-247-PPD}

\begin{abstract}

Fano noise, readout noise, and the partial charge collection (PCC) effect collectively contribute to the degradation of energy spectra in Charge Coupled Devices (CCD) measurements, especially at low energies. 
In this work, the X-ray produced by the fluorescence of fluorine (677 eV) and aluminum (1486 eV) were recorded using a Skipper-CCD, which enabled the reading noise to be reduced to 0.2~e$^-$. 
Based on an analytical description of photopeak shapes resulting from the convolution of the PCC effect and Fano noise, we achieved a precise characterization of the energy spectra. 
This description enabled us to disentangle and quantify the contributions from both Fano noise and the PCC effect. 
As a result, we determined the Fano factor and the electron-hole pair creation energy.
Additionally, we estimated the PCC-region of the sensor and, for the first time, experimentally observed the expected skewness of photopeaks at low energies.

\end{abstract}


\maketitle

\section{\label{sec:intro}Introduction}

The Fano factor $\mathcal{F}$~\cite{Fano1947} in the eV-keV range has been studied by many authors both by Monte Carlo simulations~\cite{Fraser_1994, MCCARTHY1995, MAZZIOTTA2008} and X-ray measurements~\cite{Kotov_2018, Rodrigues2021}.
The most accurate determination of $\mathcal{F}$ and electron-hole pair creation energy ($\epsilon_{eh}$) in silicon to date was performed with a Skipper-CCD at 5.9 keV using a $^{55}$Fe source~\cite{Rodrigues2021}. Measurements at lower energies are still necessary, however, in order to provide more precise values in the few eV range \cite{Durnford2018, Ramanathan2017, SENSEI2020}. 
The use of conventional CCDs introduces systematic uncertainty primarily due to readout noise ($\sigma_{RN}$), typically around $\sigma_{RN} \approx$ 1.8 e$^-$ rms/pixel for scientific CCDs~\cite{CONNIE2021}. This noise complicates the direct determination of $\mathcal{F}$, particularly at low energies where $\sigma_{RN}$ becomes significant in relation to Fano noise.
Botti et al.~\cite{Botti2022} were able to indirectly constrain $\mathcal{F}$ and $\epsilon_{eh}$ for energies below 150 eV based on Compton scattering measurements with a Skipper-CCD.  For a review of ionization modeling in silicon at low energy and further discussion of these quantities see Ref.~\onlinecite{Ramanathan2020}.

$\mathcal{F}$ in silicon is of utmost interest in particle physics since silicon-based sensors are chosen in several dark matters searches~\cite{SENSEI2020, DAMICM2020, DAMIC2021, OSCURA2022} and neutrino experiments~\cite{CONNIE2021, Moroni2020}, mainly because of their small band gap energy ($\sim$ 1.1 eV). Skipper-CCD holds the best sensitivity for light dark matter candidates~\cite{SENSEI2020} thanks to sub-electron readout noise reachable after performing multiple non-destructive measurements of the collected charge~\cite{Tiffenberg2017}.

On the other hand, an experimental technique for characterizing the effect of partial charge collection (PCC) in back-illuminated CCD was recently published~\cite{Moroni2021}. The authors have demonstrated that a backside treatment processing strongly mitigates distortions in the low energy spectra~\cite{holland2003fully}. 
However, an analytical description of the shape of energy spectra affected by this effect is still pending.
Recognizing the role of PCC in observed spectra proved to be important in order to reduce its undesired background contribution in dark matter and neutrino experiments along with optical~\cite{Pears2023}, astronomical~\cite{Marrufo2022} and cosmological applications~\cite{Drlica2020}. 

Here we report on a statistical model accounting for both Fano noise and the effect produced by PCC. As a result, we were able to accurately reproduce the experimental spectra obtained using a backside-treated Skipper-CCD. $\mathcal{F}$ and $\epsilon_{eh}$ were both determined for energies as low as 677~eV, and a first hint of skewness in the recorded spectra is observed as expected at such low energies~\cite{Fraser_1994}.

\section{\label{sec:Measurements}Measurements}

A back-illuminated fully-depleted Skipper-CCD designed by the Microsystems Laboratory at LBNL and fabricated by Teledyne DALSA Semiconductor was used for X-ray spectrometry.
The detector is divided into four quadrants, each with 443 pixels $\times$ 2063 pixels measuring 15 $\mu$m $\times$ 15 $\mu$m, and a thickness of 200 $\mu$m.
A special treatment for photon collection was performed on the backside of this sensor~\cite{holland2003fully}, which is covered by three thin layers: $\sim$20 nm indium tin oxide (ITO), $\sim$38 nm ZrO$_2$, and $\sim$100 nm SiO$_2$.
As a consequence, this detector exhibits significantly lower charge recombination compared to sensors lacking this treatment~\cite{Moroni2021}.
The detector was operated in high-vacuum conditions at 123 K using a Low-Threshold-Acquisition board~\cite{Moroni2019}.

\begin{figure}[ht]
    \centering
    \includegraphics[width=0.9\columnwidth]{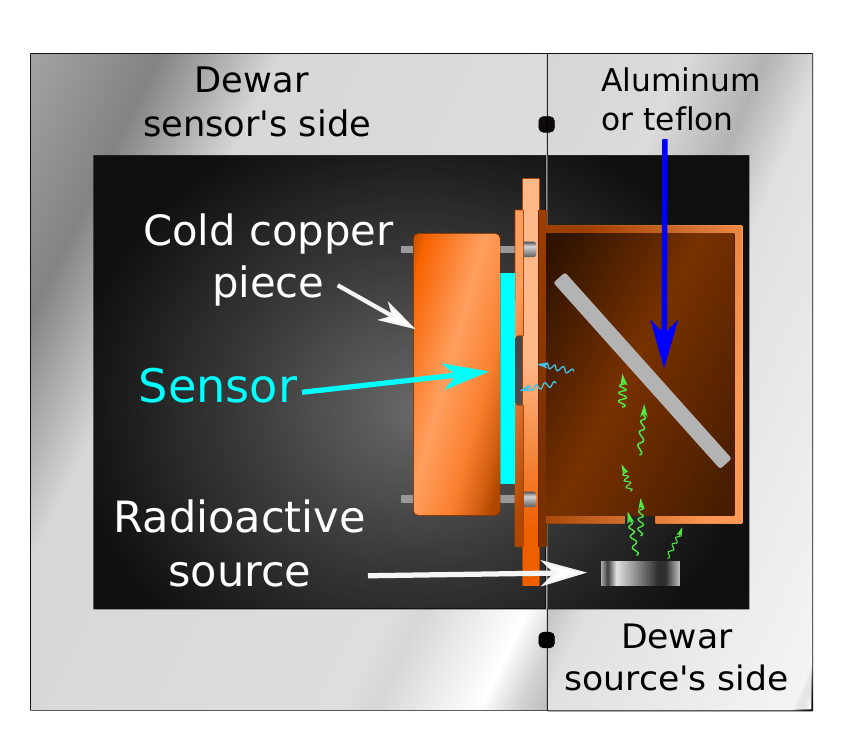}
    \caption{Experimental setup. A copper box covers the backside of the detector from which it receives X-rays produced by $\alpha$ particles hitting on surfaces with high Al or F content.}
    \label{fig:setup}
\end{figure}

Fig.~\ref{fig:setup} depicts the experimental setup. A copper box was placed in front of the detector in thermal contact with the cold plate where it was fixed. A $^{241}$Am radioactive source was situated inside the vessel just below the copper box, on which a hole was drilled to let $\alpha$ particles hit a rectangular piece of Teflon (which contains F) or Al, placed inside the box. This way, fluorescence X-rays were produced by hitting these materials. 

Fluorescence X-rays emitted by fluorine (X$_F$) have an energy of $E_F$=676.8(1)~eV, whereas those emitted by aluminum (X$_{Al}$) possess a higher energy of $E_{Al}$=1486.4(1)~eV (average of very close K-lines of 1486.70~eV and 1486.27~eV). Consequently, the attenuation length in silicon for X$_F$ is $\tau_F$=0.941(1)~$\mu$m, while for X$_{Al}$, it is $\tau_{Al}$=7.884(1)~$\mu$m~\cite{henke1993xray}.

Readout amplifiers measure the voltage produced by the charge in each pixel into Analog to Digital Units (ADUs). A self-calibration of the relationship between ADU to electron was performed in each quadrant following the same strategy described in~\cite{Rodrigues2021}. The data from each quadrant can be safely combined into a single spectrum with the aid of this absolute calibration.

In order to mitigate spatial pile-up, we limited the number of X-ray interactions in the images by reducing the exposure time, acquiring only 50 rows per quadrant instead of reading out the entire image.
For each pixel, 300 samples of the charge were averaged in order to reach a readout noise of $\sigma_{RN}$=0.2~e$^-$ for all measurements. We computed and subtracted a baseline for each row using empty pixels from an overscan zone of 50 pixels per row.
A quick cleaning procedure that takes about a second flushes out all of the charges the CCD has acquired throughout each exposure/readout cycle~\cite{holland2003fully}.

In order to avoid photons reaching the detector during readout, half of each quadrant close to the amplifiers was covered with a thin Cu foil. After exposure, the charge is moved fast below those foils, remaining shielded from X-rays until they are read.

\subsection{\label{sec:processing} Images processing}

Most of the interactions occurred within the initial roughly $10\mu$m of the backside of the Skipper-CCD employed for this research because it was back-illuminated by the X-rays from the source. 
The total number of electrons generated by each X-ray event, distributed in several pixels,  was reconstructed by running a clustering algorithm in which all non-empty neighboring pixels are grouped together.
Thanks to the sub-electron readout noise, these measurements are robust to charge transfer inefficiencies that may spread the charge between adjacent pixels since the probability of any electron from an event being separated from the other electrons by one or more empty pixels is negligible.

The charge of the events is expected to be dispersed according to a 2D Gaussian distribution due to diffusion to the surface~\cite{Haro2020}. Therefore, the standard deviation of the spatial cluster charge distribution can be used to impose a geometrical quality cut. Thus, clusters that deviate significantly from a circular geometry were suspected not to be pure, but rather to be produced by some form of pile-up.
This enables us to concentrate our study on X-ray absorption-compatible events while considering PCC-related events, as their geometry is also expected to be symmetric.

\subsection{\label{sec:background} One-electron contribution correction}

The number of pixels with only one electron in the images was significantly higher than what is observed when the Copper box and the $^{241}$Am source were not present. 
Although the main component of these events remains unknown, a possible explanation is the production of fluorescence photons emitted by Kapton in the infrared energy range due to the interaction of an $\alpha$ particle emitted by the radioactive source.
This undesired contribution eventually adds one electron to the pixels across the whole CCD with three effects as a consequence, namely: 
(a) the addition of extra charge to pixels within clusters of interest and/or, (b) the addition of charge to an empty pixel at the borders of those clusters, increasing their total charge and size and/or, 
(c) the creation of a bridge between clusters producing the agglomeration of two or more of them into a larger one. 
Effects (a) and (b) produce a bias in the amount of charge and size for a given cluster, while (c) reduces statistics due to the geometrical quality cuts applied.

\begin{figure}[ht]
    \centering
    \includegraphics[width=0.9\columnwidth]{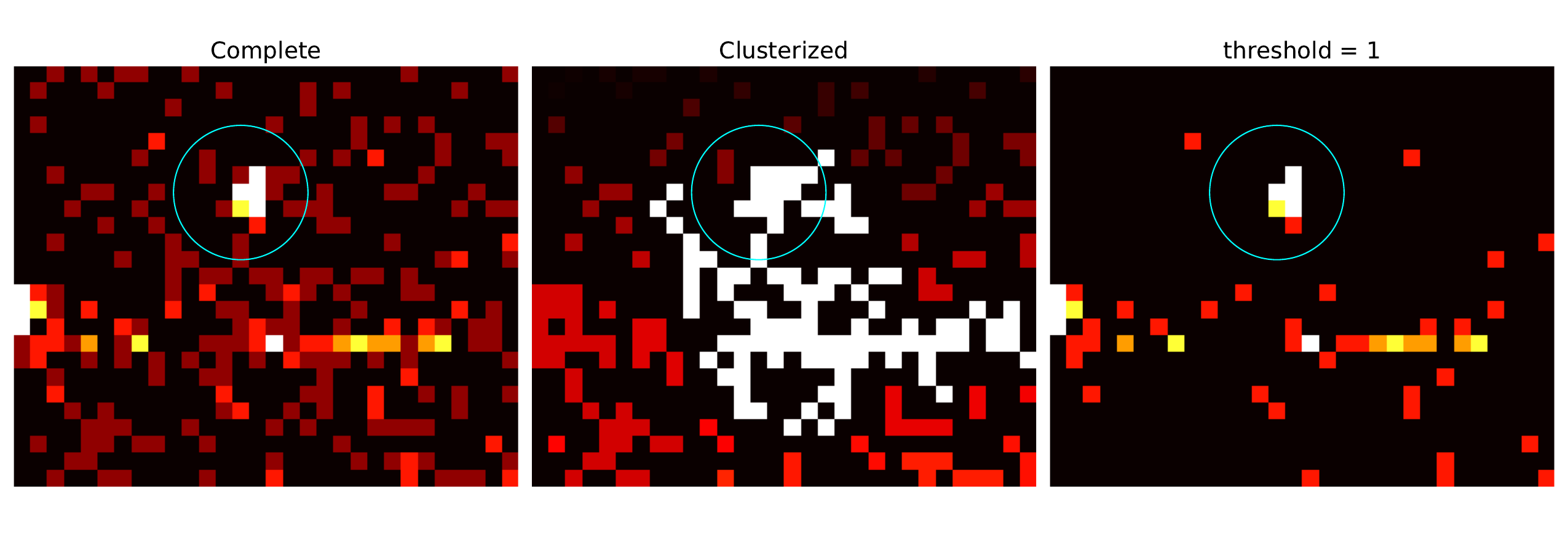}
    \caption{Left: Five-pixel cluster surrounded by one-electron pixels. Middle: After clusterization the events merged with neighbouring clusters. Right: Result after ignoring pixels with only one electron.}
    \label{fig:ClusterPegoteado}
\end{figure}

This can be seen in the leftmost image of Fig.~\ref{fig:ClusterPegoteado}, which shows a cluster of interest constituted of 6 pixels, clearly seen in white/yellow. However, after the clusterization process (middle image) is done, the resulting cluster is much larger and it is merged with a neighboring cluster. This clusterization (from leftmost to middle image) is done by defining an empty pixel as a pixel with less than 0.5 electrons. Consequently, non-empty pixels are those above 0.5 (being one-electron pixels between 0.5 and 1.5 and so on for higher occupancy levels).

In order to correct the above-mentioned undesired contribution, we redefine an empty pixel as any pixel with less than 1.5 electrons instead of 0.5. 
An example of this redefinition is shown in the rightmost image of Fig.~\ref{fig:ClusterPegoteado} that presents the reconstruction of the cluster of interest, clearly separated from its neighboring clusters.

Using a non-empty pixel threshold of 1.5 electrons solves the effects (b) and (c) previously described albeit leaving (a) unattended and creating a new undesired effect (d): the deletion of genuine one-electron pixels in the border of clusters. We refer to genuine one-electron pixels as those produced by the diffusion of events of interest to the surface, as opposed to unrelated contributions.

Undesired charge contribution can take place one, two, or more units away from clusters of interest with equal probability. While, in the case of neighboring pixels, both genuine and undesired charges will make contributions. 
Hence, a correction of effects (a) and (d) can be done \textit{ad-hoc}, event by event, by subtracting the expected undesired one-electron contribution $\mu_{und}$ and adding the one due to a genuine one-electron $\mu_{gen}$. 
In order to estimate these expectation values from the data, we measured the ratio between pixels with only one electron and empty pixels in the first and second-pixel borders of the clusters.
Those borders were determined using a pixel dilation algorithm over the clusters that contained two or more electron events per pixel. Expanding clusters one pixel in all directions and counting occurrences of one-electron pixels led to the estimation of the expected value of the sum of both contributions: $\mu_{und}$ plus $\mu_{gen}$. 
On the other hand, expanding clusters one more pixel (second border) and counting occurrences one-electron pixels, led to an estimation of $\mu_{und}$ only. Then, $\mu_{gen}$ was trivially obtained as the difference.

\begin{figure*}[t]
    \centering
    \includegraphics[width=2.0\columnwidth]{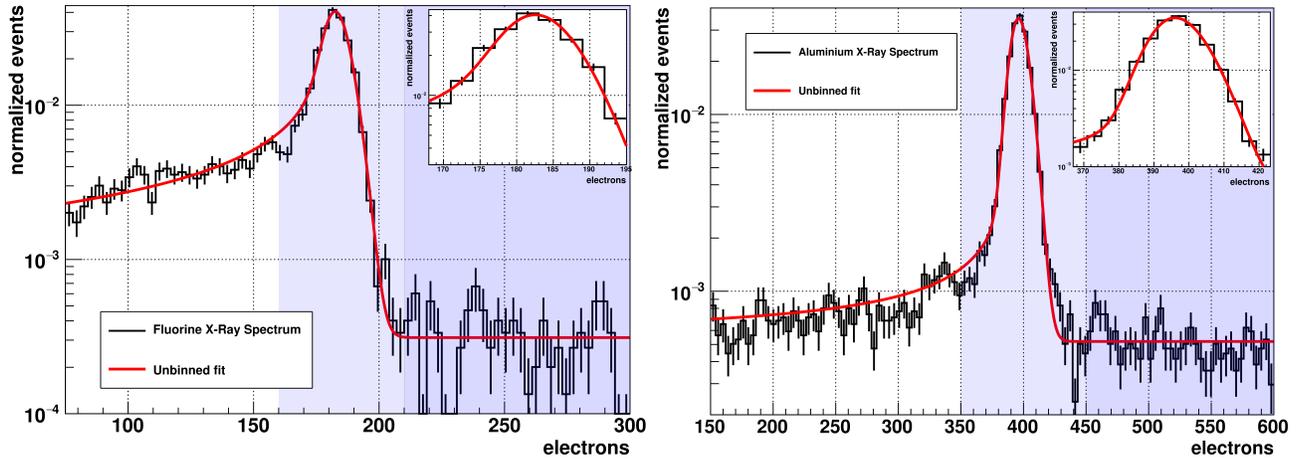}
    \caption{Normalized spectra fitted by the model from Eq.~\ref{eq:fit_function} (red line). The mean within the blue area on the right of each peak was used for background estimation. The shaded light blue areas correspond to the region not considered in the CCE calculation. The plot on the left pertains to fluorine, while the plot on the right pertains to aluminum.} 
    \label{fig:Spectra}
\end{figure*}

\begin{figure*}[ht]
    \centering
    \includegraphics[width=2.0\columnwidth]{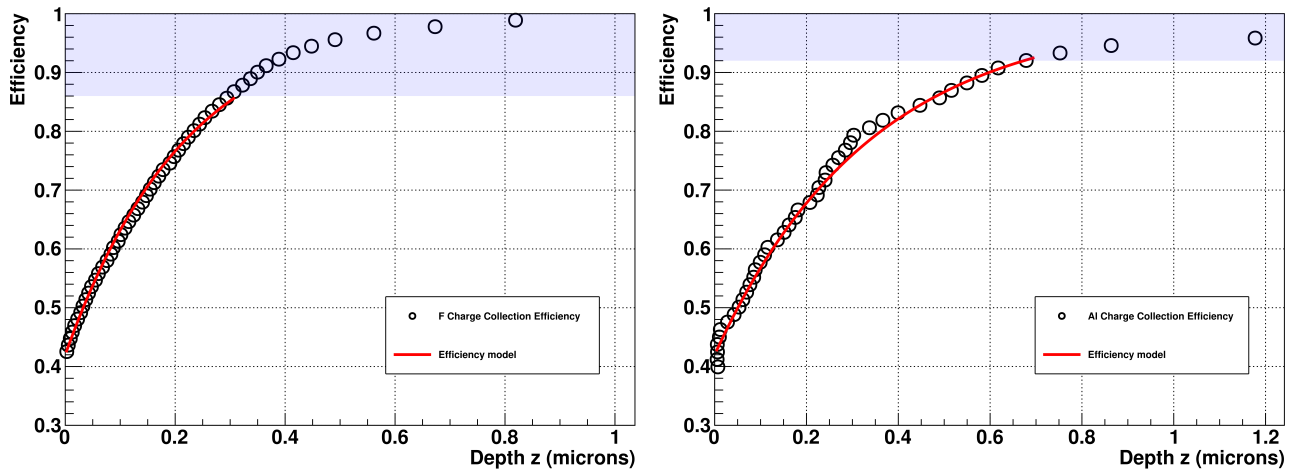}
    \caption{Charge Collection Efficiency versus sensor depth measured from the backside. The efficiency model given by Equation \ref{eq:CCE_function} is also displayed (with $\varepsilon_0$=0.42). The shaded light blue area corresponds to the $\sim$70\% of the data closer to the peak maximum not considered in the analysis since it is strongly affected by Fano noise. The plot on the left pertains to fluorine, while the plot on the right pertains to aluminum.} 
    \label{fig:PCC}
\end{figure*}

\section{X Rays spectrum and PCC}

The recorded normalized energy spectra for X$_F$ and X$_{Al}$ are presented in Fig.~\ref{fig:Spectra}, comprising 4972 and 8367 events respectively. Both exhibit a flat background to the right of the peaks, which was employed to compute the mean background per bin (blue region). Meanwhile, the left tails result from charge collection inefficiency, which characterization we will address in this subsection.

The charge collection efficiency (CCE) can be experimentally determined as a function of sensor depth using a model-independent method outlined in Ref.~\onlinecite{Moroni2021}, which can be summarized as follows.
Let $q_i$ denote the charge generated at the interaction point, \textit{i.e.} the photon energy divided $\epsilon_{eh}$ (taken as 3.75~eV~\cite{Rodrigues2021}) and let $q_f$ represent the effectively collected charge. Both $q_i$ and $q_f$ are calculated assuming no Fano noise. Additionally, consider F($q_f$) as the cumulative normalized spectra, and G($z$) as the probability that the depth of interaction is less than $z$. This approach consists of finding the value of $z$ for each q$_f$ such that G($z$) = F($q_f$). Subsequently, for each ($z$, $q_f$) pair, the efficiency function is calculated as 
\begin{equation}
\varepsilon(z)=q_f/q_i.
\label{eq:qfqi}
\end{equation}

The depth $Z$ into the detector at which an X-ray interacts with the CCD is a random variable that follows an exponential distribution, which probability density function (PDF) is described by
\begin{equation}
g_Z(z)=\frac{1}{\tau_X}\exp\Bigl(-\frac{z}{\tau_X}\Bigr)
\label{eq:exponential}
\end{equation}
where $\tau_{X}$ represents the attenuation length. Therefore, G($z$) can be calculated as the integral of Eq.~\ref{eq:exponential} between 0 and $z$.
Due to $\tau_{Al} > \tau_F$, X$_{Al}$ penetrate deeper into the silicon material compared to X$_F$. Consequently, the former is less influenced by the PCC region as can be seen in Fig.~\ref{fig:Spectra}. 
The results of applying this technique are presented in Fig.~\ref{fig:PCC}.

In this work, in order to analytically model $\varepsilon(z)$, we employ the following function:
\begin{equation}
\varepsilon(z)=1-(1-\varepsilon_0)\exp\Bigl(-\frac{z}{\tau_{CEE}}\Bigr)
\label{eq:CCE_function}
\end{equation}
The parameter $\tau_{CEE}$ is introduced to characterize the size of the PCC-region, while $\varepsilon_0$ represents the efficiency at $z=0$ (the backside illuminated by X-rays). 
Due to the backside treatment of the sensor used in this study, we expect $\varepsilon_0$ not to be zero. 
This may be perceived as a step in the low-energy range, situated where the tail caused by PCC effect concludes, which originates from the peak. 
However, the observation of this step is affected by Fano noise, resulting in blurring. This Fano effect was used in Ref.~\onlinecite{Botti2023} to constrain the Fano factor at 150 eV measuring Compton steps at the L-shell energies.

Fig.~\ref{fig:PCC} shows how the function given by Eq.~\ref{eq:CCE_function} (red line) reasonably described both observed dependence for X$_F$ and X$_{Al}$. Interestingly, they yield compatible $\varepsilon_0$ values around 0.42, however, $\tau_{CEE}$ differs a $\sim$50\%. Such difference could be explained by to no geometrical corrections have been made to account for variations in the arrival direction of X-rays originating from a surface as large as the detector (see Fig.~\ref{fig:setup}). Such a correction is described in Ref.~\cite{Moroni2021} and will be treated in detail for this data in Ref.~\cite{Botti2023}. 
To mitigate the impact of the Fano effect on CCE calculation, approximately 70\% of the data points closer to the peak maximum were excluded from the analysis. This corresponds to the light blue zone in both Fig.~\ref{fig:Spectra} and Fig.~\ref{fig:PCC}.

\section{Spectra Shape Model}
With the intention of deriving an analytical expression for the PDF of the number of charges measured in each event, we will first find the PDF for $\varepsilon(z)$ and then convolve it with the one accounting for Fano noise.

The randomness of $\varepsilon(z)$ is caused by the randomness of $z$. Therefore, in order to find its PDF, $f_{E}(\varepsilon)$, we use Eq.~\eqref{eq:exponential}, \eqref{eq:CCE_function} and the transformation rule for random variables:
\begin{equation*}
f_{E}(t)=g_Z(\varepsilon^{-1}(t))\Biggl\lvert\frac{d\varepsilon^{-1}(t)}{dt}\Biggr\rvert
\end{equation*}
\noindent to get
\begin{equation*}
f_{E}(\varepsilon)=\frac{\beta}{1-\varepsilon_0} \Bigl(\frac{1-\varepsilon}{1-\varepsilon_0}\Bigr)^{\beta-1} \quad 
\end{equation*}

\noindent where $\beta$ corresponds to the ratio between the $\tau$'s defined above, that is
\begin{equation}
\beta=\frac{\tau_{CCE}}{\tau_X}
\label{eq:beta}
\end{equation}

\noindent Note that for $\varepsilon_0$=0, $f_{E}(\varepsilon)$ becomes a Beta distribution with $\alpha$=1.

On the other hand, in the absence of the effect of the PCC, and for relatively high energies, the PDF for the charge produced by X-rays, $Q_X$, is expected to be described by a Gaussian, $N(q_X|\mu, \sigma^2)$ where $\mu$ is the mean and $\sigma^2$ the variance.
Montecarlo simulations predict, however, an increasing skewness, \textit{i.e.} an asymmetry PDF, as the energy decreases \cite{Fraser_1994}. 
We use a skew-gaussian distribution to model this effect, which is:
\begin{equation*}
f_{Q_X}(q_X|\mu, \sigma, \lambda)= 2 N(q_X|\mu, \sigma) \Phi\Bigl[\lambda \frac{q_X-\mu}{\sigma}\Bigr]
\end{equation*}
\noindent with 
\begin{equation*}
\Phi(\lambda y)=\frac{1}{2} \Bigl[1+Erf\Bigl(\frac{\lambda y}{\sqrt{2}}\Bigr)\Bigr]
\end{equation*}
\noindent where $Erf(.)$ is the error function and $\lambda$ a parameter that allows to modify the skewness~\cite{Azzalini_Capitanio}. 
For details about how $\lambda$ also modifies the mean value $\langle Q_X \rangle$, and the variance $Var(Q_X)$ see Appendix~\ref{sec:appendix}.

Combining all the aforementioned information, we are now able to calculate the joint probability.
\begin{equation*}
f_{E \times Q_X}(\varepsilon,q_X)=2 N(q_X|\mu, \sigma) \Phi\left[\frac{\lambda}{\sigma} \Bigl(q_X-\mu \Bigr)\right]
\end{equation*}
\noindent and following a variable change ($q_X=q_f/\varepsilon$), we integrate to obtain the desired PDF.
\begin{equation}
\label{eq:fit_function}
f_{Q_f}(q_f) = \int_0^1 \frac{2}{\varepsilon} N\left(\frac{q_f}{\varepsilon}|\mu, \sigma\right) \Phi\left[\frac{\lambda}{\sigma} \Bigl(q_f-\mu \Bigr)\right] f_{E}(\varepsilon) d\varepsilon 
\end{equation}
Note that in the previous variable substitution, $q_X$ replaces $q_i$ in the CCE calculation from Eq.\ref{eq:qfqi}, as $q_X$ now represents the initial charge while incorporating fluctuations due to Fano noise.

Finally, a minor contribution from a uniform distribution $U_{Q_f}$ is introduced to address the background events observed on the right side of each peak, which is assumed to persist as flat as is observed throughout the entire analyzed range.
The background events likely result from the Compton scattering of more energetic photons interacting with the CCD. Since the probability of Compton scattering caused by fluorine or aluminum fluorescence X-rays is completely negligible compared to photoelectric absorption, we do not apply an extra correction due to this effect.

\section{\label{sec:Results} Results and discussion}

A likelihood unbinned fit based on the PDF from Eq.~\ref{eq:fit_function} plus $U_{Q_f}$ was performed in order to get the best set of estimated parameters ($\hat{\mu}$, $\hat{\sigma}$, $\hat{\beta}$, $\tau_{CCE}$, $\hat{\lambda}$) was calculated from $\hat{\beta}$ using Eq.~\ref{eq:beta}.
The Skewness was calculated from them. The relative weight of $U_{Q_f}$ was estimated as the ratio between the flat rate of events observed in the blue region in Fig.~\ref{fig:Spectra} propagated over the full analyzed range and the total number of events. The PDF for the best set of parameters is plotted in Fig.~\ref{fig:Spectra} with a red line.

Table~\ref{tab:table1} summarises the results for all the quantities of interest in this work. $\mathcal{F}$ and $\epsilon_{eh}$ were calculated from the fitted parameters and the energy $E_X$ of X-ray photons by means of these equations:
\begin{equation}
\mathcal{F}=\frac{Var(Q_f)}{\langle Q_f \rangle}\quad \text{and} \quad
\epsilon_{eh}=\frac{E_X}{\langle Q_f \rangle}
\end{equation}

\begin{table*}
\renewcommand{\arraystretch}{1.5}
\caption{\label{tab:table1} Mean number $\langle Q_f \rangle$, standard deviation $\sqrt{Var(Q_f)}$ and Skewness from both X$_{F}$ and X$_{Al}$ spectra. Effective size of the PCC-region $\tau_{CCE}$, Fano factor $\mathcal{F}$ and electron-hole creation energy $\epsilon_{eh}$ calculated from them and the energy of X$_{F}$ and X$_{Al}$.}
\begin{ruledtabular}
\begin{tabular}{ccccccc}
  X-ray source&$\langle Q_f \rangle$&$\sqrt{Var(Q_f)}$&Skewness&$\tau_{CCE}[nm]$&$\mathcal{F}$&$\epsilon_{eh}[eV]$\\ \hline
 F &  180.8(2) & 6.8(1) & 0.26(7) & $\sim$230 & 0.137(6) & 3.748(2) \\ \hline
 Al & 392.9(4) & 10.2(2) & 0.20(4) & $\sim$450 & 0.146(9) & 3.783(4) \\ \hline
\end{tabular}
\end{ruledtabular}
\end{table*}

The spectra seem to exhibit a subtle indication of the anticipated low-energy step accounting for no zero $\varepsilon_0$ in the leftmost bins. Nevertheless, it is important to mention that all the results presented in this work are resilient against this parameter. Even for null $\varepsilon_0$, the change in the reported results is within their uncertainties.

\paragraph {Impact of the sub-electron readout noise}

To assess the impact of the sub-electron readout noise per pixel ($\sigma_{RN}$ = 0.2 e$^-$), we calculate the resulting noise within each cluster. The distribution of cluster sizes conforms well to a Poisson distribution, with an expectation value of 10.1 for X$_{F}$ and 12.4 for X$_{Al}$. Conversely, the variance of each cluster is derived from the summation of variances across individual pixels, all of which are read out with the same read noise. This leads to an average of $\sim$0.7 electrons per cluster. In contrast, the same calculation for a standard scientific CCD yields a total of $\sim$7 electrons. 

\begin{figure}[t]
    \centering
    \includegraphics[width=1.0\columnwidth]{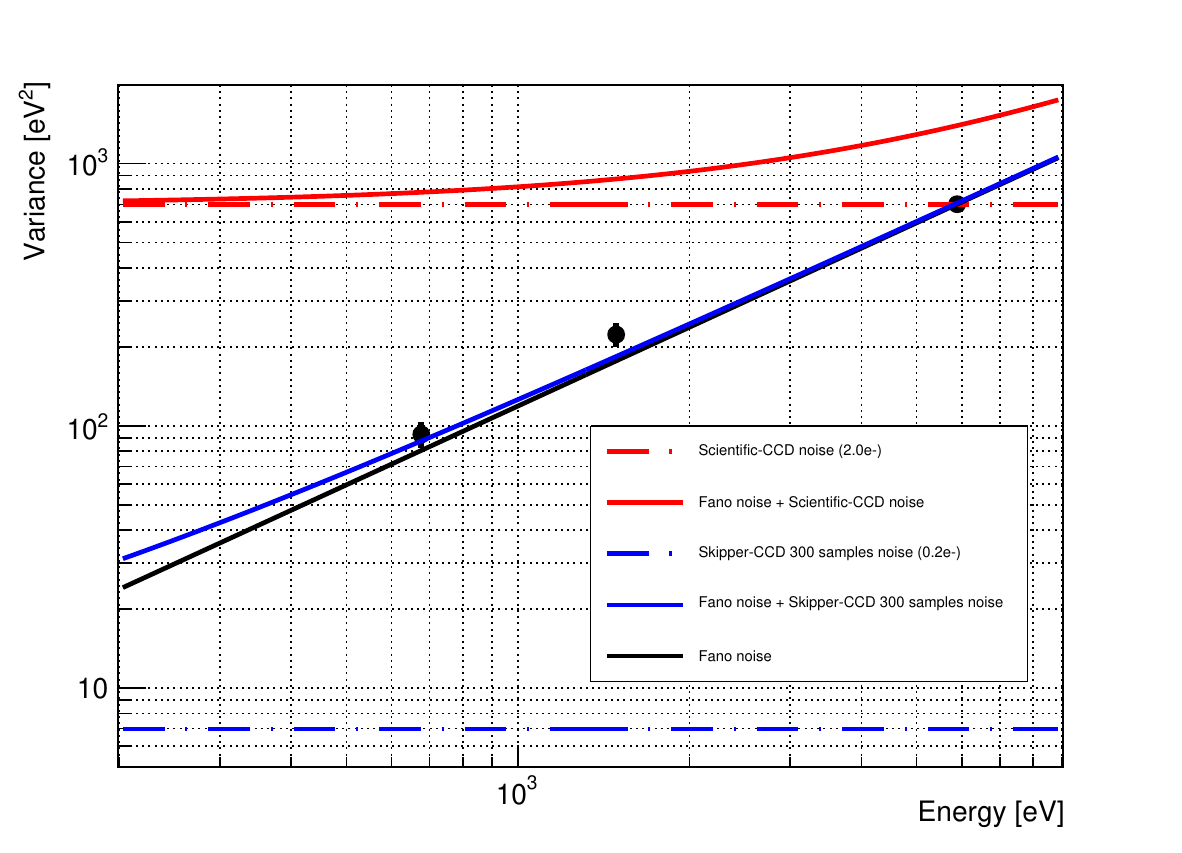}
    \caption{Variance $Var(Q_f)$, scales with $\varepsilon_{eh}^2$, as a function of the mean energy $\langle Q_f \rangle \times \varepsilon_{eh}$ for the charge distribution obtained for X$_F$ and X$_{Al}$. Additionally, we have included the value for $^{55}$Fe X-rays ($\sim$5.9 keV) from~\cite{Rodrigues2021}. The horizontal lines on the graph represent the readout noise levels for a conventional scientific CCD and a Skipper-CCD after conducting 300 samples per pixel. The black line signifies the expected value for a Fano factor of 0.119, irrespective of the energy.} 
    \label{fig:VarvsE}
\end{figure}

Figure~\ref{fig:VarvsE} exemplifies the benefits of employing a Skipper-CCD for this type of measurement. The curves depict various scenarios: the black line represents the expected values in the absence of any additional noise (pure Fano noise). 
Furthermore, it presents the anticipated results for measurements conducted using a standard scientific CCD, indicated by the red lines, alongside those obtained with a Skipper-CCD after 300 samples, following the methodology used in this study and denoted by the blue line. 
The figure highlights the advantage of employing Skipper-CCD technology for such measurements, enabling the determination of the Fano factor in an energy range where it would otherwise be dominated by the readout noise of other technologies.

\paragraph {Statistical uncertainties}
In order to compute the statistical uncertainties, the Likelihood function was marginalized for each fitted parameter. 
The 68\% CL intervals were determined as the range where each marginalized LogLikelihood was greater than its maximum minus 0.5.
The uncertainty on $\mathcal{F}$, $\epsilon_{eh}$ and Skewness were computed by means of a Monte Carlo propagation throughout its mathematical relationship with the fitted parameters (see Appendix~\ref{sec:appendix}). 

\paragraph {Systematic uncertainties} The resilience of the results against energy ranges was investigated. As a result, the small changes observed in the best set of estimated parameters preserve the results reported in Table \ref{tab:table1} within their confidence intervals. 
It is also worth mentioning that when $\mathcal{F}$ and $\epsilon_{eh}$ are calculated considering null skewness, the results still lie within the reported confidence interval. Thus, although a hint of skewness is observed in the data, its inclusion in the model has a negligible impact on $\mathcal{F}$, $\epsilon_{eh}$, and its uncertainties.

Systematic contributions from quality cuts were carefully computed and found to be negligible when compared to the statistical uncertainty, with the exception of $\Delta \langle Q_f \rangle$.
Because of its very low statistical uncertainty ($\sim$0.1\%), the systematic contribution to $\Delta \langle Q_f \rangle$ becomes significant and was added in quadrature.

Regarding $\tau_{CCE}$'s uncertainty, it is important to note that the model assumes the photons to have normal incidence, albeit this condition was not satisfied during the experiment. As it can be seen from Fig.~\ref{fig:setup}, the angle of incidence of each photon depends on both the point over the material surface where the fluorescence originated and the point on the sensor surface where the incident photon interacts. Thus, a difference in the position of the materials used as targets for $\alpha$ particles to hit results in a different effective path through the silicon. 
For all of this, just an estimated value of $\tau$ is reported in Table \ref{tab:table1}. A detailed study of the latter effect is reserved for a future work.

The influence of other potential factors, such as misclassification during clustering, charge transfer inefficiencies, and charge loss in the skippering process can be confidently disregarded in this study. 
This confidence is based on previous investigations employing the same sensor and image processing, which estimated their combined systematic contribution negligible for $^{55}$Fe measurements~\cite{Rodrigues2021}. 
By contrast, in this study, the statistical uncertainties are three times higher than those obtained in that work, making the systematic contribution even more insignificant in comparison.

$\mathcal{F}$ for both F and Al are compatible within uncertainties, however, they turn incompatible with the previous results at 5.9 keV, which is $\mathcal{F}$=0.119(2)~\cite{Rodrigues2021}. This result seems to indicate that this quantity increases as the energy decreases even though further measurements at even lower energies are necessary to confirm this hypothesis.

Regarding $\epsilon_{eh}$, the result obtained from $X_F$ is compatible with the previously reported at 5.9 keV of 3.752(2) eV~\cite{Rodrigues2021}, however, the one from $X_{Al}$ is incompatible within their uncertainties although being just 0.8\% higher.

\section{\label{sec:Conclusions} Conclusions}

A statistical model to describe the shape of an X-ray spectrum acquired by CCD was derived based on the convolution of Fano noise and the effect of the PCC-region. This model was used to fit the photopeak produced by fluorescence X-ray spectra at 677~eV and 1486~eV. 
As a result, a hint of skewness was observed at 677~eV as expected by Monte Carlo simulations.

The same model provides an estimate for the effective size of the PCC region. We observed that a minimum of 90\% of the charge generated by photons penetrating beyond approximately 600 nm is collected.
This value falls within the same range as previous measurements using a different technique and X-ray source~\cite{Moroni2021}, reaffirming that the PCC region can be significantly reduced through backside processing.
It is worth noting that the developed model enables this estimation by utilizing information from the spectrum, even in proximity to the photopeak, while also accounting for the possible presence of a minimal non-zero efficiency.
The PCC-region would be determined more precisely with this model if additional measurements ensuring normal X-ray incidence are performed.

In summary, through the use of a Skipper-CCD in conjunction with an analytical spectral shape model, we determined $\mathcal{F}$ and $\epsilon_{eh}$ below 1 keV, where, unlike in other technologies, readout noise does not dominate. These results remain robust against the effects of partial charge collection, which were disentangled by the analytical model.

\begin{acknowledgments}
This work was supported by Fermilab under DOE Contract No.\ DE-AC02-07CH11359. 
This manuscript has been authored by Fermi Research Alliance, LLC under Contract No. DE-AC02-07CH11359 with the U.S.~Department of Energy, Office of Science, Office of High Energy Physics. 
The CCD development work was supported in part by the Director, Office of Science, of the U.S. Department of Energy under Contract No. DE-AC02-05CH11231. 
The United States Government retains and the publisher, by accepting the article for publication, acknowledges that the United States Government retains a non-exclusive, paid-up, irrevocable, world-wide license to publish or reproduce the published form of this manuscript, or allow others to do so, for United States Government purposes.
SU was supported in part by the Zuckerman STEM Leadership Program and DR by the National Council for Scientific and Technical Research (CONICET). DR acknowledges the support of Agencia Nacional de Promoción de la Investigación, el Desarrollo Tecnológico y la Innovación through grant PICT 2018-02153.
\end{acknowledgments}

\appendix
\section{Skew-Normal Density Function}
\label{sec:appendix}
Following the prescriptions established in the work of Alzzani~\cite{Azzalini_Capitanio}, a Skew-Normal Density Function can be obtained as
\begin{equation*}
f_X(x|\mu, \sigma, \lambda)= N(x|\mu, \sigma) \Bigl\{1+Erf\Bigl[\frac{\lambda}{\sqrt{2}} \Bigl(\frac{x-\mu}{\sigma}\Bigr)\Bigr]\Bigr\}
\end{equation*}
where $N(x|\mu, \sigma)$ is the PDF for a Gaussian density function center at $\mu$ with a variance of $\sigma^2$, and
with
\begin{equation*}
Erf(y)=\frac{2}{\pi}\int_0^y e^{t^2} dt
\end{equation*}\

The mean value, $\langle X \rangle$, and the Variance, $Var(X)$ depends on $\lambda$ as follows:
\begin{eqnarray*}
\langle X \rangle= \mu + \sigma b \rho \quad \quad Var(X) = \sigma^2 \Bigl[1 - b^2 \rho^2 \Bigr]
\label{eq:EandVar}
\end{eqnarray*}
with
\begin{eqnarray*}
\rho= \frac{\lambda}{\sqrt{1+\lambda^2}} \quad and \quad b=\sqrt{\frac{2}{\pi}} 
\label{eq:rhoandb}
\end{eqnarray*}
and the Skewness is given by 
\begin{eqnarray*}
\gamma(X)=&\frac{4-\pi}{2} \Biggl[\frac{(\langle X \rangle-\mu)^2}{Var(X)}   \Biggr]^{3/2}
\label{eq:skewness}
\end{eqnarray*}

\nocite{*}

\bibliography{apssamp}

\end{document}